\documentclass{article}
\usepackage{graphicx}
\usepackage{hyperref}
\begin{document}
\newcommand{\pl}[1]{Phys. Lett. {\bf #1}}
\newcommand{\zpb}[1]{Zeit. Phys. Lett. {\bf #1}}
\newcommand{\prl}[1]{ Phys. Rev. Lett. {\bf #1}}
\newcommand{\pr}[1]{Phys. Rev. {\bf #1}}
\newcommand{\apb}[1]{Ann. of Phys. (N.Y.) {\bf #1} }
\newcommand{\np}[1]{Nucl. Phys. {\bf #1}}

\begin{flushright} 

IFUP-TH 44/00 \end{flushright}
\vskip 35pt plus 3pt minus 3pt

\begin{center}
  {\large {\bf Coulomb-Nuclear Coupling  and  Interference Effects in the  Breakup of Halo
Nuclei} } 
\end{center}
\vspace{1em}
\begin{center}
{Jerome Margueron}\footnote 
{\small Electronic address : margueron@ganil.fr}

{\it Istitut de Physique Nucl\'eaire , IN2P3-CNRS, 91406  Orsay, France.}\footnote {Present address GANIL, bp:
5027;14076 CEAN CEDEX 5, France. }\end{center}

\begin{center}
{Angela Bonaccorso}\footnote 
{\small Electronic address : angela.bonaccorso@pi.infn.it}

{\it Istituto Nazionale di Fisica Nucleare,
 Sezione di Pisa, 56100 Pisa, Italy,}\end {center}

\begin{center}
{David M. Brink}\footnote 
{\small Electronic address : brink@ect.it}

{\it Department of Theoretical Physics, 1 Keble Road, Oxford OX1 3NP, U. K.}

\end{center} 

\date{today}

\begin{center}
{\bf Abstract}
\end{center}
Nuclear and Coulomb breakup of halo nuclei have been treated  often as 
incoherent processes and structure information have been extracted from their study. The
aim of this paper is to clarify whether interference effects and Coulomb-nuclear
couplings are important and how they could modify the simple picture previously used.
 We calculate the neutron angular and energy distributions by using first order
perturbation theory for the Coulomb amplitude and an eikonal approach for the
nuclear breakup. This allows for a simple physical interpretation of the
results which are mostly analytical. Our formalism includes the effect of the
nuclear distortion of the neutron wave function on the Coulomb amplitude. This
leads to a Coulomb-nuclear coupling term derived here for the first time which
gives a small contribution for light targets but is of the same order of
magnitude as nuclear breakup for heavy targets. The
overall interference is  constructive for light to medium targets and destructive for heavy
targets. Thus it appears that Coulomb breakup experiments need to be analyzed with more
accurate models than those used so far.

\begin{flushleft} 
 {\bf PACS }
number(s):25.70.Hi, 21.10Gv,25.60Ge,25.70Mn,27.20+n
 \end{flushleft}

\begin{flushleft}  {\bf Key-words}
Nuclear breakup, Coulomb breakup, interference, halo nuclei.\end{flushleft}
\section{Introduction}

This paper is concerned with the nuclear and Coulomb breakup of weakly bound nuclei, in
particular halo nuclei in which the last neutron is very weakly bound and in a state of low
angular momentum. As a consequence its distance from the core center is quite large and this
makes the nuclear and Coulomb breakup   strong effects in particular when the target is heavy.

There have been a large number of approaches \cite{tb}-\cite{wa} attempting to give an
accurate description of this phenomenon and the most recent papers contain a review of the
present situation
\cite {tb,ts,he}.
The accuracy of the method is an important issue not only from the point of view of
the understanding of the reaction mechanism but also because  a number
of  parameters of the neutron initial state wave function such as the binding energy and the
spectroscopic factor can be deduced from the data analysis 
\cite{tb}-\cite{me}. 

 An important
aspect of the problem is that nuclear breakup is always present at the same time. It is
dominant for a light target and it remains non negligible for heavy targets. Its
contribution to  the total breakup cross section on a heavy target has been a matter of
debate but it should range from about 20\% to 50\%. In such conditions one expects  to see
interference effects. However most of the analysis of experimental
 data so far performed have
relied on the assumption that interference effects are small in the intermediate energy range (40-80 A.MeV).
Furthermore nuclear and Coulomb breakup have been treated as independent processes and
calculated in the absence of each other to different orders in the interactions. Such
approximations have been justified by the fact that the nuclear interaction is strong and
has a short range, therefore needs to be treated to all orders. The Coulomb potential on
the other hand is comparatively weaker and has a long range, therefore it has often been
treated to first order.

 One difficulty in the
theoretical approach comes from the fact that one needs to have a consistent description
of the nuclear and Coulomb breakup amplitudes from which the interference originates. In
this respect the most accurate calculations presently available are the numerical
solutions of the Schr\"odinger equation \cite{ts}. They treat both interactions to all
orders and therefore contain also all possible interference effects. However it is not so
easy to extract from them a simple picture of the coupling mechanisms. The numerical
solution of the Schr\"odinger equation has also been used with only the nuclear
\cite{eb1} or the Coulomb potential \cite{tf,kys,mb} to test the accuracy of simpler
models used to calculate the nuclear and Coulomb breakup separately. The conclusions of
recent works have been that, for well developed haloes, nuclear breakup needs to be
treated to all orders but the sudden and eikonal approximations are acceptable since they
give an accuracy of about 20\%
\cite{eb1,BB1} compatible with the present experimental error bars. On the other hand it
seems that Coulomb breakup can be safely treated to first order \cite{tb,ts,tf}, but the
time dependence needs to be retained in order to reproduce some very fine experimental
observables such as the neutron-core final energy spectrum. DWBA \cite{ban,cha,ban1},
semiclassical \cite{dlv,dlv1,jas},  adiabatic  \cite{trj} and coupled channel
\cite{nt} approaches have also been proposed. A common feature of some of those
approaches
\cite{ts,mb,dlv1,jas} has been  to give a smaller Coulomb breakup contribution than that 
calculated with the first order perturbation theory.  Also many of them were concerned
with proton breakup \cite{nt,eh} and lower energy regimes. Such situations are
physically more challenging and we do not attempt to include them in the present
discussion. Therefore we restrict ourselves to neutron breakup at intermediate energies
in the attempt to clarify the role of Coulomb-nuclear couplings and interference and to
make a link between numerical solutions and simple analytical methods. 

 An important issue is to determine 
 the experimental observable on which couplings and interference effects would show
up best. Interferences are a quantum mechanical manifestation of a coherence effect.
Therefore in order to see them an observable of the "exclusive" type is expected to be
most suitable.  Some work has been done
\cite{dlv,nt,esbe,st} to study interference effects on the ejectile angular distribution
but nothing exists at present on the interference effects on the neutron angular
distribution which we address here.  Neutron angular distributions have been the best
experimental proof of the existence of the two distinct mechanisms of Coulomb and nuclear diffractive breakup.
Furthermore they give information on the relative strength of the two processes, on the target dependence and  beam
energy dependence of the reaction mechanisms.

Neutron angular distributions due to nuclear and Coulomb breakup were measured and
calculated in
\cite{anne} but information on the interferences were not extracted. 
Another reliable calculation of the neutron angular distribution from nuclear breakup
alone was given in
\cite{ab}.  Models for the reaction mechanism and structure information extracted
from the data can be considered reliable only if  three aspects of the data are reproduced at the same
time within a given model. One is the ejectile or neutron energy spectrum or its equivalent
parallel momentum distribution.  Another is the neutron angular distribution. Finally the absolute breakup  cross
section is important to fix the initial state parameters.

In this paper we will try to get some insight on the problem by a
simple but consistent calculation of the Coulomb and  the nuclear breakup in the presence
of each other and of their interference.  This will allow us to get both the  neutron
angular distribution as well as its energy spectrum due to the combined effects of the
nuclear and Coulomb interaction. The absolute breakup cross  section is obtained from the
integration of the angular distribution. We will discuss under which experimental  conditions
interference effects could be seen and why they have not been clearly identified so far. 
An important result of our method is that it gives simple analytical expressions for the
Coulomb cross section  for any initial angular momentum state of the neutron.

Section 2 of the paper presents some theoretical background. The purpose is to
clarify the assumptions made in deriving the nuclear and Coulomb breakup
amplitudes. Section 3 gives expressions for the cross-section and the breakup
amplitudes. The results of calculations are presented in section 4 and
conclusions in section 5. Some details of the derivation of the Coulomb
amplitude are given in the first Appendix. The calculations presented in the
paper assume an s-state initial neutron wave function in the projectile. Some
formulae which generalize the results for the Coulomb amplitude are given in
the second appendix. They are included for completeness and are not used in
the present paper.

\section{Theoretical background}

We consider the breakup of a halo nucleus like $^{11}$Be consisting of a
neutron bound to a $^{10}$Be core in a collision with a target nucleus. In
previous works on halo breakup \cite{ab,bbn} we have assumed that the projectile
core and the target move along classical trajectories. Here we adopt a
different approach. The system of the halo nucleus and the target is described
by Jacobi coordinates $\left(  \mathbf{R,r}\right)  $ where $\mathbf{R}$ is
the position of the center of mass of the halo nucleus relative to the target
nucleus and $\mathbf{r}$ is the position of the neutron relative to the halo
core, and the coordinate $\mathbf{R}$ is assumed to move on a classical path.
This allows target recoil to be included in a consistent way. The Hamiltonian of the system is%

\begin{equation}
H=T_{R}+T_{r}+V_{nc}\left(  \mathbf{r}\right)  +V_{2}\left(  \mathbf{R,r}%
\right)  \label{h1}%
\end{equation}
where $T_{R}$ and $T_{r}$ are the kinetic energy operators associated with the
coordinates $\mathbf{R}$ and $\mathbf{r}$ and $V_{nc}$ is a real potential
describing the interaction of the neutron with the core. The potential $V_{2}$
describes the interaction between the projectile and the target. It is a sum
of two parts depending on the relative coordinates of the neutron and the
target and of the core and the target%
\begin{equation}
V_{2}\left(  \mathbf{R,r}\right)  =V_{nt}\left(  \mathbf{\beta}_{2}%
\mathbf{r+R}\right)  +V_{ct}\left(  \mathbf{R-\beta}_{1}\mathbf{r}\right)
\label{pot1}%
\end{equation}
Here $\beta_{1}=m_{n}/m_{p}$, $\beta_{2}=m_{c}/m_{p}=1-\beta_{1}$, where
$m_{n}$ is the neutron mass, $m_{c}$ is the mass of the projectile core and
$m_{p}=m_{n}+m_{c}$ is the projectile mass. Both $V_{nt}$ and $V_{ct}$ are
represented by complex optical potentials. The imaginary part of $V_{nt}$
describes absorption of the neutron by the target to form a compound nucleus.
It gives rise to the stripping part of the halo breakup. The imaginary part of
$V_{ct}$ describes reactions of the halo core with the target. The potential
$V_{ct}$ also includes the Coulomb interaction between the halo core and the
target. This part of $V_{ct}$ is responsible for Coulomb breakup.

The mass ratio $\beta_{1}$ is small for a halo nucleus with a heavy core. For
example $\beta_{1}\approx0.1$ and $\beta_{2}\approx0.9$ in the case of $^{11}%
$Be. This property is used here to approximate the neutron-target and
neutron-core potentials by

\begin{eqnarray}
V_{nt}\left(  \mathbf{\beta}_{2}\mathbf{r+R}\right)&  \approx&
V_{nt}(\mathbf{r+R)}\label{vnt}\\
V_{ct}\left(  \mathbf{R-\beta}_{1}\mathbf{r}\right)& \approx&
V_{ct}(\mathbf{R})\mathbf{+V}_{eff}\left(  \mathbf{r,R}\right)  \label{vnc}%
\end{eqnarray}

\bigskip where
\begin{equation}
\mathbf{V}_{eff}\left(  \mathbf{r,R}\right)  =\mathbf{\beta}_{1}%
\mathbf{r\cdot F}_{ct}\left(  \mathbf{R}\right)  \qquad\mathrm{and}\qquad
\mathbf{F}_{ct}\left(  \mathbf{R}\right)  =-\nabla V_{ct}\left(
\mathbf{R}\right)  \label{veff} \end{equation}

Here \
$\mathbf{F}_{ct}\left(  \mathbf{R}\right)  $ is the classical force
acting between the target and the projectile core. The halo breakup is caused
by the direct neutron target interaction $V_{nt}$ or by a recoil effect due to
 the core-target interaction. Coulomb breakup of a one-neutron halo nucleus
is a recoil effect due the Coulomb component $V_{ct}$ of the core-target
interaction and is contained in $\mathbf{V}_{eff}\left(  \mathbf{r,R}\right)$. It is
proportional to the mass ratio $\beta_{1}$. 
The dipole expansion Eqs.(\ref{vnc}),(\ref{veff}) is a good approximation provided $\beta
_{1}r<<R$. In the case considered here $r$ is of the order of the halo radius
(4 to 6 fm) and $R$ is larger than the core-target strong absorption radius (
about 6 fm for a $^{9}$Be target and 12 fm for a Pb target). As $\beta
_{1}\approx0.1$ \ for the $^{11}$Be halo nucleus the dipole approximation
should be valid.

The nuclear part of the core-target interaction can also contribute to
$\mathbf{V}_{eff}$ and gives rise to the 'shake-off' component of the breakup
amplitude which is discussed in ref.\cite{bh}. They show that it
can be important for a light target. It is not included in the present paper. 

The theory in this paper is based on a time dependent approach which can be
derived from an eikonal approximation. The projectile motion relative to the
target is described by a time-dependent classical trajectory $\mathbf{R}%
\left(  t\right)  =\mathbf{d}+vt\mathbf{\hat{z}}$ with constant velocity $v$
and impact parameter $\mathbf{d}$ ( $\mathbf{\hat{z}}$ is a unit vector
parallel to the z-axis). With the approximations (\ref{vnt},\ref{vnc}) and
(\ref{veff}) to the potentials the wave function $\phi\left(  \mathbf{r,d}%
,t\right)  $ describing the dynamics of the halo neutron satisfies the
time-dependent equation
\begin{equation}
i\hbar\frac{\partial\phi\left(  \mathbf{r,}\mathbf{d},t\right)  }{\partial
t}=\left(  H_{r}+V_{nt}(\mathbf{r+R}\left(  t\right)  \mathbf{)+V}%
_{eff}\left(  \mathbf{r,R}\left(  t\right)  \right)  \right)  \phi\left(
\mathbf{r,}\mathbf{d},t\right)  \label{tde}%
\end{equation}
where $H_{r}=T_{r}+V_{nc}\left(  \mathbf{r}\right)  $ is the Hamiltonian for
the halo nucleus. As $t\rightarrow-\infty$ the wave function tends to the
initial halo nucleus wave-function
\begin{equation}
\phi\left(  \mathbf{r,}\mathbf{d},t\right)  \rightarrow\phi_{lm}\left(
\mathbf{r},t\right)  =\phi_{lm}\left(  \mathbf{r}\right)  \exp\left(
-i\varepsilon_{0}t/\hbar\right)  \label{wf0}%
\end{equation}
with binding energy $\varepsilon_{0}$. The initial state $\phi_{lm}\left(
\mathbf{r}\right)  $ can be an s-state or a general single particle state with
angular momentum $\left(  l,m\right)  $.

In the present paper we neglect the final state interactions between the
neutron and the halo core, but include the final state interactions between
the neutron and the target. This approximation should be satisfactory unless
there are resonances in the neutron-core final state interaction which give rise to modification in the momentum
distributions of the fragments of the type recently discussed in \cite{eb1}. Such effects
have not been seen so far in the experimental distributions. When the neutron-core final
state interactions are neglected the breakup amplitude can be written as%
\begin{equation}
g_{lm}\left(  \mathbf{k,}\mathbf{d}\right)  =\frac{1}{i\hbar}\int_{-\infty
}^{\infty}dt\left\langle \phi_{f}\left(  t\right)  |\bar{V}_{2}\left(
\mathbf{r},t\right)  |\phi_{lm}\left(  t\right)  \right\rangle
\end{equation}
where $\bar{V}_{2}\left(  \mathbf{r,}t\right)  =V_{nt}(\mathbf{r+R}\left(
t\right)  \mathbf{)+V}_{eff}\left(  \mathbf{r,R}\left(  t\right)  \right)  $
and $\phi_{f}\left(  t\right)  $ satisfies the equation%
\begin{equation}
i\hbar\frac{\partial\phi_{f}\left(  t\right)  }{\partial t}=(T_{r}+\bar{V}%
_{2}\left(  \mathbf{r},t\right)  )\phi_{f}\left(  t\right)
\end{equation}
with the boundary condition that $\phi_{f}\left(  t\right)  \sim\exp\left(
i\mathbf{k \cdot r}-i\varepsilon_{\mathbf{k}}t/\hbar\right)  $ when $t$ is large. The
final step is to make an eikonal approximation for $\phi_{f}\left(  t\right)
$%
\begin{equation}
\phi_{f}\left(  t\right)  =\exp\left(  i\mathbf{k \cdot r}-i\varepsilon_{\mathbf{k}%
}t/\hbar\right)  \exp\left(  -\frac{1}{i\hbar}\int_{t}^{\infty}\bar{V}%
_{2}\left(  \mathbf{r},t^{\prime}\right)  dt^{\prime}\right)  \label{wfe}%
\end{equation}
The breakup amplitude becomes%
\begin{equation}
g_{lm}\left(  \mathbf{k,}\mathbf{d}\right)  =\frac{1}{i\hbar}\int
d^{3}\mathbf{r}\int dte^{-i\mathbf{k \cdot r}+i\omega t}e^{\left(  \frac{1}{i\hbar
}\int_{t}^{\infty}\bar{V}_{2}\left(  \mathbf{r},t^{\prime}\right)  dt^{\prime}\right)  }\bar
{V}_{2}\left(  \mathbf{r},t\right)  \phi_{lm}\left(  \mathbf{r}\right)
\label{amp1}%
\end{equation}
where $\omega=\left(  \varepsilon_{\mathbf{k}}-\varepsilon_{0}\right)  /\hbar
$. \ 

The components $V_{nt}$ and $V_{eff}$ of $\bar{V}_{2}\left(  \mathbf{r}%
,t\right)  $ have to be treated differently because of the long range of the
Coulomb interaction in $V_{eff}$. The neutron-target interaction is strong and
has a short range. In this paper we assume that the interaction time $\tau
_{f}$ for this part of the interaction is very short in the sense that
$\omega\tau_{f}$ is small compared with unity. On the other hand the dominant
contribution to $V_{eff}$ comes from the long range Coulomb interaction
between the halo core and the target. It is weaker and changes more slowly. In
this paper we calculate the contribution of $V_{eff}$ to first order but
include its full time dependence. We will show that under such hypothesis the total
amplitude
$g_{lm}\left(
\mathbf{k,}\mathbf{d}\right)  $ can be written as a sum of two parts
\begin{equation}
g_{lm}\left(  \mathbf{k,}\mathbf{d}\right)  =g_{lm}^{nucl}\left(
\mathbf{k,}\mathbf{d}\right)+g_{lm}^{Coul}\left(  \mathbf{k,}\mathbf{d}%
\right)
\end{equation}
where $g_{lm}^{nucl}\left(  \mathbf{k,}\mathbf{d}\right)  $ is zero order in
$V_{eff}$ and $g_{lm}^{Coul}\left(  \mathbf{k,}\mathbf{d}\right)  $ is first
order. Explicit expressions for the Coulomb and nuclear contributions \ are
given in the next section.

\section{Amplitudes and Cross Sections}
 
\subsection{The nuclear amplitude}

   The nuclear breakup amplitude is obtained from eq.(\ref{amp1}) by expanding to first
order in $V_{eff}$ and separating the term which is zero order in 
$V_{eff}$. We assume that the interaction time is so short that the $\omega
$-dependence in eq.(\ref{amp1}) can be neglected. This is the sudden
approximation or 'frozen halo approximation'. The integral can be evaluated by
changing the time variable to $t^{\prime}=t+z/v$.  The nuclear breakup amplitude
 reduces to%

\begin{eqnarray}
g_{lm}^{nucl}\left(  \mathbf{k,}\mathbf{d}\right)   &  =&\int d^{3}\mathbf{r}
e^{-i\mathbf{k.r}}(\exp\left(  -i\chi_{nt}\left(  \mathbf{b}\right)\right)
-1  )\phi_{lm}\left(  \mathbf{r}\right) \label{amp3}\\&  = &\int
d^{2}\mathbf{r}_{\perp}e^{-i\mathbf{k}_{\perp}\mathbf{.r}_{\perp} }
\left(\exp\left(  -i\chi_{nt}\left( 
\mathbf{b}\right)\right)-1    \right)
\tilde\phi_{lm}\left(  \mathbf{r}_{\perp},k_{z}\right) \nonumber
\end{eqnarray}
\bigskip where the eikonal phase%
\begin{equation}
\chi_{nt}\left(  \mathbf{b}\right)  =\frac{1}{\hbar}\int_{-\infty}^{\infty
}V_{nt}\left(  \mathbf{r}_{\perp}\mathbf{+d},t\right)  dt. \label{ph3}%
\end{equation}
\bigskip The vector $\mathbf{b=r}_{\perp}+\mathbf{d}$ is the impact parameter
of the neutron relative to the target. The breakup amplitude (\ref{amp3}) is
equivalent to the one used in \cite{ab,BB1} except that in those references the
dependence of the neutron-target interaction on the combination $z+vt$ is
taken into account before setting $\omega=0$. The effect is to replace the
z-component $k_{z}$ of the final momentum by $k_{1}$ where%

\begin{equation}
k_{1}=k_{z}+\frac{\omega}{v}=k_{z}+\frac{\varepsilon_{\mathbf{k}}%
-\varepsilon_{0}}{\hbar v}.\end{equation}

\bigskip

\subsection{The Coulomb amplitude}

The calculation of Coulomb breakup is different from nuclear breakup because:
{i)} The Coulomb force has a long range. {ii)} The Coulomb force does not act
directly on the neutron but it affects it only indirectly by causing the
recoil of the charged core. The recoil effect is included in the effective
interaction $V_{eff}$. The Coulomb contribution can be obtained by putting the
Coulomb potential in the definition (\ref{veff}) and is%

\begin{eqnarray}
V_{eff}(\mathbf{r,R}(t))  &  =&+{\beta}_{1}Z_{P}Z_{T}e^{2}\left(
\frac{\mathbf{r\cdot R}\left(  t\right)  }{R\left(  t\right)  ^{3}}\right)
\label{veff2}\\
&  =&+{\beta}_{1}Z_{P}Z_{T}e^{2}\frac{xd+zvt}{\left(  d^{2}+\left(  vt\right)
^{2}\right)  ^{3/2}} \label{veff3}%
\end{eqnarray}
Equation (\ref{veff2}) is the effective Coulomb potential on the neutron in
the dipole approximation. The effective force on the neutron (\ref{veff}) is
in the opposite direction to $\mathbf{R}\left(  t\right)  $.

The fact that the Coulomb interaction has a long range means that the
dependence of the breakup amplitude on the excitation energy $\varepsilon
_{\mathbf{k}}-\varepsilon_{0}=\hbar\omega$ cannot be neglected. In the present
section we give an expression for the breakup amplitude which is correct to
first order in the Coulomb interaction and which also contains the effects of
the neutron-target interaction as well as the $\omega$-dependence. The Coulomb
contribution to the breakup amplitude $g_{lm}^{Coul}\left(  \mathbf{k,}%
\mathbf{d}\right)  $ is obtained from (\ref{amp1}) by writing $\bar{V}%
_{2}=V_{nt}+V_{eff}$ and expanding  to first order in $V_{eff}$. The
first order contribution to the integrand in (\ref{amp1}) is proportional to%
\begin {eqnarray}
&  e^{-i\chi_{nt}\left(  \mathbf{r,}t\right)  }\left(  V_{eff}\left(
\mathbf{r},t\right)  -i\chi_{eff}\left(  \mathbf{r,}t\right)  V_{nt}\left(
\mathbf{r},t\right)  \right) \nonumber\\
&  =V_{eff}\left(  \mathbf{r},t\right)  -\left(  1-e^{-i\chi_{nt}\left(
\mathbf{r,}t\right)  }\right)  V_{eff}\left(  \mathbf{r},t\right)
-ie^{-i\chi_{nt}\left(  \mathbf{r,}t\right)  }\chi_{eff}\left(  \mathbf{r,}%
t\right)  V_{nt}\left(  \mathbf{r},t\right)  \label{eamp2}%
\end{eqnarray}
where%
\begin{equation}
\chi_{nt}\left(  \mathbf{r,}t\right)  =\frac{1}{\hbar}\int_{t}^{\infty}%
V_{nt}\left(  \mathbf{r},t^{\prime}\right)  dt^{\prime},\qquad\chi_{eff}\left(  \mathbf{r,}%
t\right)  =\frac{1}{\hbar}\int_{t}^{\infty}V_{eff}\left(  \mathbf{r},t^{\prime}\right)
dt^{\prime}
\end{equation}

The qualitative effects of the neutron-target interaction in the Coulomb breakup amplitude
for any value of $\omega$ can be understood by making a simple approximation
to the nuclear part of the eikonal phase in eq.(\ref{eamp2});%
\begin{equation}
\chi_{nt}(\mathbf{r,}t)=\chi_{nt}\left(  \mathbf{b}\right)  \theta\left(
t\right)  ,\qquad\qquad\label{ph4}%
\end{equation}
where $\theta\left(  t\right)  =0,t>0;\quad\theta\left(  t\right)  =1,t<0$.
\ The physical assumption is that the neutron target interaction is important
only for a short time around $t=0$ corresponding to the point of closest
approach between the projectile and the target. The Coulomb effective
interaction is more slowly varying with time and has a contribution for $t>0$
when the nuclear phase eq.(\ref{ph4}) is zero and a contribution for $t<0$
when the nuclear phase is important. With this approximation the Coulomb
amplitude can be written as a sum of two terms%
\begin{equation}
g_{lm}^{Coul}\left(  \mathbf{k,}\mathbf{d}\right)  =g_{lm}^{pert}\left(
\mathbf{k,}\mathbf{d}\right)  +g_{lm}^{nC}\left(  \mathbf{k,}\mathbf{d}%
\right)  \label{amp4}%
\end{equation}

The first term is obtained from the first term in eq.(\ref{eamp2}). It depends
only on the Coulomb interaction and yields the standard perturbation result
for Coulomb breakup in the dipole approximation%
\begin{equation}
g_{lm}^{pert}\left(  \mathbf{k,}\mathbf{d}\right)  =-i\int d^{3}\mathbf{r}%
e^{-i\mathbf{k.r}}\chi_{eff}\left(  \mathbf{r},\omega\right)  \phi_{lm}\left(
\mathbf{r}\right)  \label{amp5}%
\end{equation}
where%
\begin{equation}
\chi_{eff}\left(  \mathbf{r},\omega\right)  =\frac{1}{\hbar}\int_{-\infty
}^{\infty}dte^{i\omega t}V_{eff}\left(  \mathbf{r},t\right)
\end{equation}

The second part of (\ref{amp4}) has contributions from the second and the
third terms in eq.(\ref{eamp2}). In the third term we treat again the neutron-target interaction in the sudden
approximation. Then the sum of these two contributions is%
\begin{equation}
g_{lm}^{nC}\left(  \mathbf{k,}\mathbf{d}\right)  =\int d^{3}\mathbf{r}%
e^{-i\mathbf{k.r}}(1-\exp\left(  -i\chi_{nt}\left(  \mathbf{b}\right)
\right)  )B\left(  x,z,\varpi\right)  \phi_{lm}\left(  \mathbf{r}\right)
\label{amp6}%
\end{equation}
where%
\begin{equation}
B\left(  x,z,\varpi\right)  =\frac{i}{\hbar}\left[  \int_{-\infty}%
^{0}dte^{i\omega t}V_{eff}\left(  \mathbf{r},t\right)  +\int_{0}^{\infty
}dtV_{eff}\left(  \mathbf{r},t\right)  \right]  \label{amp7}%
\end{equation}
Explicit expressions for\ $g_{r}^{pert}\left(  \mathbf{k,}\mathbf{d}\right)
$\ and $B\left(  x,z,\varpi\right)  $\ are given in the appendix. The
amplitude (\ref{amp6}) has an interesting structure. It is exactly like the
nuclear breakup amplitude (\ref{amp3}) but with an effective neutron wave
function%
\begin{equation}
\phi_{eff}\left(  \mathbf{r,\varpi}\right)  =B\left(  x,z,\varpi\right)
\phi_{0}\left(  \mathbf{r}\right)
\end{equation}
If $\phi_{0}\left(  \mathbf{r}\right)  $ is an s-state then $\phi_{eff}\left(
\mathbf{r,\varpi}\right)  $ is a p-state.

The results presented in the next section have been obtained with the previous formulae,
 where the
sudden approximation or frozen halo approximation has  been made only for
the neutron-target interaction. The same approximation can also
be made for the Coulomb amplitude by putting the Coulomb adiabaticity
parameter $\varpi=\omega d/v$ equal to zero in the above
expressions. When    $\hbar\omega=\varepsilon_{\mathbf{k}}-\varepsilon_{0}%
\approx1$ MeV, $E_{lab}=41$ A.MeV and $d\approx R_{s}=11.5$ fm then
$\varpi\approx0.10$. This number is relatively small so the sudden
approximation might also be reasonable for the Coulomb amplitude in some cases. Thus the
expression for the complete amplitude (\ref{amp1}) in the sudden approximation
becomes%
\begin{equation}
g_{lm}\left(  \mathbf{k,}\mathbf{d}\right)  =\int d^{2}\mathbf{r}_{\perp
}e^{-i\mathbf{k}_{\perp}\mathbf{.r}_{\perp}}\left(  \exp\left(  -i(\chi
_{nt}\left(  \mathbf{b}\right)  +\chi_{eff}\left(  \mathbf{b}\right)
)\right) -1 \right) \tilde \phi_{lm}\left(  \mathbf{r}_{\perp},k_{z}\right)
\label{sudd}\end{equation}
The Coulomb part to first order is%
\begin{equation}
g_{lm}^{Coul}\left(  \mathbf{k,}\mathbf{d}\right)  =-i\int d^{2}
\mathbf{r}_{\perp}e^{-i\mathbf{k}_{\perp}\cdot \mathbf{r}_{\perp}}\chi_{eff}\left(
\mathbf{b}\right)  \exp\left(  -i\chi_{nt}\left(  \mathbf{b}\right)\right) \tilde \phi_{lm}\left( 
\mathbf{r}_{\perp},k_{z}\right) \end{equation}
which is consistent with eqs.(\ref{amp4},\ref{amp5}) and (\ref{amp7}).

\subsection{\bf Cross sections}\label{cb}
This section contains the formulas needed for calculating neutron breakup cross sections.
The final state for the 3-body system in the breakup reaction is specified by
the momentum $\mathbf{k}$ conjugate to coordinate $\mathbf{r}$\ and the
momentum $\mathbf{K}$\ conjugate to $\mathbf{R}$. They are related to final
momenta of the core, neutron and target by
\[
\mathbf{k}_{c}=-\mathbf{k+\beta}_{2}\mathbf{K,\qquad k}_{n}=\mathbf{k}%
+\beta_{1}\mathbf{K,\qquad k}_{T}=-\mathbf{K.}
\]
In the eikonal approximation for the core-target scattering the full 3-body breakup
amplitude
$A\left( 
\mathbf{K,k}\right) 
$ is given by%
\begin{equation}
A\left(  \mathbf{K,k}\right)  =\int d^{2}\mathbf{d}e^{-i\mathbf{K}_{\perp
}\mathbf{.d}}S_{ct}\left(  \mathbf{d}\right)  g\left(  \mathbf{k,d}\right)
\label{sca}%
\end{equation}
where $S_{ct}\left(  \mathbf{d}\right)  $\ is the profile function describing
the scattering of the halo core by the target. It is given in terms of the
core-target potential by the eikonal formula%
\begin{equation}
S_{ct}\left(  \mathbf{d}\right)  =\exp\left(  -\frac{i}{\hbar}\int_{-\infty
}^{\infty}dtV_{ct}\left(  \mathbf{d},vt\right)  \right)  .\label{pf}%
\end{equation}
The differential cross-section integrated over $\mathbf{K}_{\perp}$ is \
\[
\frac{d\sigma}{d^{3}\mathbf{k}}=\frac{1}{\left(  2\pi\right)  ^{5}}\int
d^{2}\mathbf{K}_{\perp}|A\left(  \mathbf{K,k}\right)  |^{2}=\frac{1}{\left(
2\pi\right)  ^{3}}\int d^{2}\mathbf{d|}S_{ct}\left(  \mathbf{d}\right)
|^{2}|g\left(  \mathbf{k,d}\right)  |^{2}.
\]
If $\beta_{1}$ is small then $\mathbf{k}_{n}$ is almost the same as
$\mathbf{k}$.  If ${\bf k}_f=(k_x,k_y,k_{f_z})$ is the neutron final momentum in the 
target reference frame,
while 
${\bf k}=(k_x,k_y,k_z)$ is in the projectile reference frame and $k_z=k_{f_z}-m_nv/\hbar$,
$k_\perp^2=k_x^2+k_y^2$, then the
above cross section is equivalent to the expression for
\ the cross section for emitting neutrons with energy $\varepsilon_{f}$ into
the solid angle $d\Omega$. It is obtained by integrating the breakup probability
distribution over the impact parameter $\mathbf{d}$ of the center of mass of
the projectile relative to the target \cite{bb}

\begin{equation}{\frac{d^{2}\sigma}{d\varepsilon_{f}d\Omega}}=C^{2}S\int_{0}^{\infty}
d^{2}\mathbf{d} {d^2P({\bf
k_f,d})\over d\varepsilon_fd\Omega}P_{el}(d),\label{csec}\end{equation} where
$P_{el}=|S_{ct}\left( 
\mathbf{d}\right)  |^{2}$ is the probability that the projectile core-target system
remains in the ground state during the reaction. Several simple parameterizations are
possible for $P_{el}$. The profile function $S_{ct}\left(  \mathbf{d}\right)  $ can be
calculated from the core-target optical potential with the eikonal formula (\ref{pf}). It
can also be approximated by the strong absorption model
\begin{eqnarray} P_{el}=1 \,\,\,\hbox{ for }\,\,  d>R_s \nonumber \\ P_{el}=0\,\,\,
\hbox{ for }\,\, d<R_s\label{prs1}\end{eqnarray} 
where $R_{s}$ is the strong absorption radius for the core-target collision.
 In this way the calculations are analytical up to the cross
section formula \cite{ab}. The breakup probability is obtained from%

\begin{equation}
{\frac{d^{2}P(d)}{d\varepsilon_{f}d\Omega}}={\frac{1}{8\pi^{3}}}%
{\frac{m_nk_{f}}{\hbar^{2}}}{\frac{1}{2l+1}}\sum_{m}|A_{lm}^{nucl}+A_{lm}%
^{Coul}|^{2}.\label{anc}%
\end{equation}
The initial neutron bound state wave function in the projectile has angular
momentum $\left(  l,m\right)  $ and the amplitudes $A_{lm}^{nucl}$ and
$A_{lm}^{Coul}$ are related to $g_{lm}\left(  \mathbf{k,}\mathbf{d}\right)  $
by a phase factor%
\begin{equation}
A_{lm}^{nucl}+A_{lm}^{Coul}=e^{-i\mathbf{k\cdot d}}g_{lm}\left(  \mathbf{k,}%
\mathbf{d}\right)  .\label{amp2}%
\end{equation}

The relation between the neutron-core relative energy and the neutron energy in the
laboratory is
\begin{equation} \varepsilon_{\mathbf{k}}=\varepsilon_f+{1\over 2}m_nv^2-k_{f_z}v\hbar.
\end{equation}

Eq.(\ref{amp2}) makes a connection with the notation in earlier works \cite{bb}.
The phase factor has its origin in a different choice  of the coordinate
system. 

In eq.(\ref{anc}) the momentum of the breakup neutron has polar angles
$\left(  \theta,\phi\right)  $ where the azimuthal angle $\phi$ refers to the
theoretical reaction plane defined by the incoming beam momentum
$\mathbf{k}_{in}$ and the impact parameter $\mathbf{d}$. This plane is not an
observable but when $K_{\perp}d>>1$ then  it is almost the same as the plane
defined by the incoming direction and the direction of $\mathbf{K=k}%
_{c}+\mathbf{k}_{n}$. This is because the main contribution to the integral
(\ref{sca})\ for $A\left(  \mathbf{K,k}\right)  $\ comes from the range of
angles where $\mathbf{K}_{\perp}$ is almost parallel to or almost antiparallel to
$\mathbf{d}$.

\section {Results}

In this section we discuss  some sample results for the breakup of the halo state
in $^{11}Be$ obtained by using the formalism of the previous sections,  and  in order to
asses the accuracy of the formalism we will compare to existing experimental data.
Initial state parameters are the same as in \cite{ab}. Namely the initial neutron binding energy for the 2s state is
$\varepsilon_0=-0.5MeV$ and the spectroscopic factor $C^2S=0.77$ \cite{as}. The neutron-target optical
potential, defined as in \cite{67},
 for the Be target is given in Table 1 and it is the same as \cite{bbn} for the other targets. We will show
separately the cross sections due only to the nuclear amplitude, to the Coulomb
amplitude and their coherent and incoherent sums. The
Coulomb amplitude eq.(\ref{amp4}) is a sum of two parts: the perturbation amplitude
eq.(\ref{amp5}) which depends only on the Coulomb interaction and the Coulomb-nuclear
term eq.(\ref{amp6}) which contains the distorting effect of the nuclear interaction
on Coulomb breakup. Both neutron energy spectra and
angular distributions will be discussed. 

\begin{table}
   \caption{ Energy dependent optical model parameters for $n+^{9}$Be.
  a$_R$=0.387fm, r$_I$=1.368fm, a$_I$=0.3fm at all    energies.
$R_R=(1.447A_T^{1/3}-0.005(\varepsilon_f-20))fm.$}
 \begin{center}
   \begin{tabular}{|cccc|} \hline
   $\varepsilon_f$&$V_R$&$4 W_S$&
   $W_V$\\
   (MeV)&(MeV)&(MeV)&(MeV)\\\hline
   20-40&38.5-0.145$\varepsilon_f$&
   1.666+0.365$\varepsilon_f$&0.375$\varepsilon_f-7.5$\\ 
   40-120&&16.226-0.1($\varepsilon_f-40)$&7.5-0.02($\varepsilon_f-40)$\\ 
   120-180&&8.226-0.07($\varepsilon_f-120)$&5.9\\ 
          \hline
   \end{tabular}\end{center}\end{table}

The main purpose of the calculations presented here is  first to clarify
some differences in the mechanisms of Coulomb and nuclear breakup, then to asses the
importance of the Coulomb-nuclear coupling term and of the interference effects. Their
influence on the final neutron energy distribution is discussed in section 4.1 and on the
energy-integrated angular distribution in section 4.2. The first results for the angular
distribution in section 4.2 are integrated over the azimuthal angle of the outgoing
neutron direction. This corresponds to an experimental situation where the final
direction of the halo core is not measured. 

In the second part of the discussion in section 4.2 the angular distributions
are integrated over a limited range of azimuthal angles measured with respect
to the plane of the classical trajectory. As explained at the end of the first
part of section 3 this plane is almost the same as the plane defined by the
incoming direction and the direction of $\mathbf{K=k}_{c}+\mathbf{k}_{n}$
provided $\left|  \mathbf{K}_{\perp}\right|  $ is large enough. In any case a
measurement of such an angular distribution requires a measurement of both
$\mathbf{k}_{n}$ and $\mathbf{k}_{c}$.

\subsection{Energy distributions}
\begin{figure}[ht]
\centering
\includegraphics[scale=0.5,angle=-90]{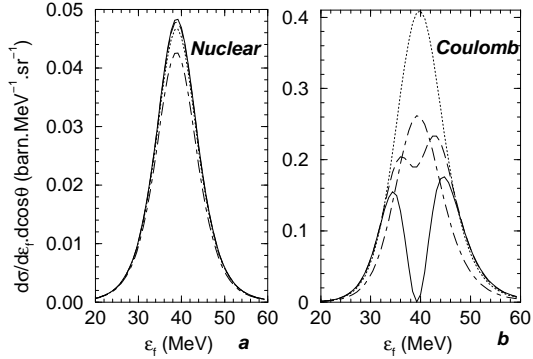}
\caption{Neutron energy distributions following nuclear (a) and Coulomb (b) breakup of $^{11}$Be at 41A.MeV at fixed
$\theta$ ($^{197}$Au target).  Solide line $\theta=0$ degree, dashed line  $\theta=1.5$ degrees, 
doted line  $\theta=3$ degrees, dashed-dot line  $\theta=6$ degrees.}
\end{figure}

In Figs. (1a) and (1b), we have represented the nuclear and Coulomb breakup spectra  for
different values of the  angle
$\theta$ as indicated, in the case of the Au target. In Fig. (1b) the Coulomb cross
section at very forward angles ( $\theta=0$) is zero   for
$\epsilon_f\approx m_nv^2/2$ and it has then two asymmetric peaks.
 The double hump structure in the Coulomb breakup cross-section is a known feature of dipole Coulomb breakup
from DWBA (eg. calculations by Okamura \cite{ok}). However it was stated in \cite{ok} that the origin of the double
peak was not clear in the DWBA formalism. Its origin is instead very clear in our semiclassical treatment.

In fact from the approximation used in  Eq.(\ref{vnc})  one sees that the core target interaction
gives rise to an effective interaction on the neutron Eq. (\ref{veff}). Then  as Coulomb breakup is a reaction which
transfers some momentum to the neutron, there cannot  be breakup when the incident neutron keeps both its incident
velocity and direction.  This kinematical effect can be easily understood by looking at Eqs.(\ref{amp5}) and
(\ref{amp8}). The initial wave function momentum distribution has a maximum at $k_z=0$ for s and p states. In
such conditions
$k_{f_z}=k_f cos \theta={m_nv\over \hbar}$ and the neutron going at zero degrees should keep all the
 momentum available from the relative motion. In fact the
Coulomb operator contains a derivative with respect to $k_z$ which then gives a zero
probability for such a process.  On the other hand if the neutron is  slightly deflected it
can keep his incident energy with a maximum probability. The peaks at the other angles
correspond to kinematical situations where the neutrons have all the available incident
energy per nucleon at the distance of closest approach d corresponding to the velocity 
$v=(2(E_{CM}-V_{CB}(d))/\mu )^{1/2} $ where $\mu$ is the projectile-target reduced mass. Such an
energy is slightly less than the incident energy per particle because of the Coulomb barrier.
This is always the case for nuclear breakup as shown in Fig. (1a). The kinematics of
Coulomb breakup can be easily understood by looking at Fig. (2) where we show the relevant
momentum vectors in the laboratory reference frames, as indicated. The incident momentum
per particle is
$m{\bf v}$ and
 ${\bf p^{\prime}}$, ${\bf p^{\prime\prime}}$ are the two neutron momenta in the
projectile reference frame corresponding to the same given neutron  laboratory angle
$\theta$. $|{\bf p^{\prime}}|$ and $|{\bf p^{\prime\prime}}|$ are equal to the
averaged momentum transferred to the neutron by the Coulomb effective field. Along the
laboratory final momentum axis we sketch the corresponding energy distribution as
already given in Fig. (1b). It is clear that increasing the angle $\theta$ the two peaks
tend to join each other, merging finally to just one peak at a critical  angle 
$\theta_c=\sqrt{2m V_{CB}/A_P}/ \mu v$ 
beyond which Coulomb breakup is kinematically suppressed. Note
that $\theta_c$ is the angle at which the Coulomb breakup angular distribution is a
maximum, as shown later in Fig.(5).

\begin{figure}[ht]
\centering
\includegraphics[scale=0.4,angle=0]{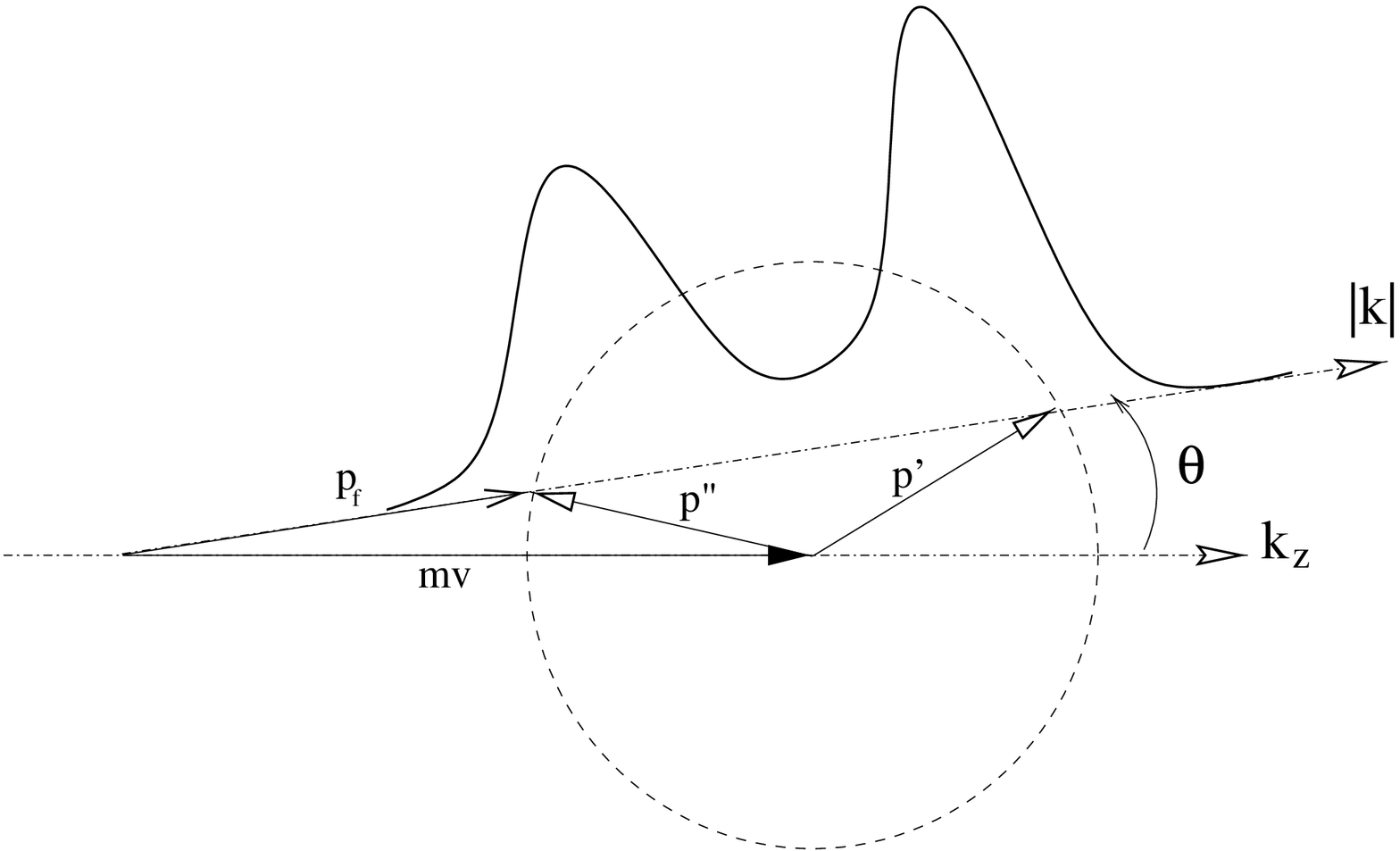}
\caption{Kinematics of the breakup reaction. The solid curve represents the
Coulomb energy distribution for a given small $\theta$.}
\end{figure}

On the other hand, in the nuclear breakup, Fig.(1a), the neutron average momentum in the
projectile reference frame is $k_z \approx 0$, corresponding to the maximum of the 
initial momentum distribution and of the breakup probability.
Then the energy distribution has just one peak at all angles, close to the incident 
energy (and momentum) per particle.

Double peaked longitudinal distributions are typical of reactions in which the mechanism
of breakup involves an effective repulsive force between the fragments. For example in heavy-ion
fragmentation reactions the so called Coulomb rings are observed \cite{cha1}. This happens
because the reaction goes through a kind of inelastic excitation in the projectile which then
decays in flight.  Therefore it is suggested that exclusive measurements of longitudinal
distributions with the  neutron in coincidence with the core, for fixed, small, angles $\theta$
could help distinguishing different reaction mechanisms.  This could be very useful in the case
of two-neutron halo breakup in which the second neutron decays in flight \cite{bro,6he,norr} from a
resonant state and therefore with a mechanism different from the nuclear breakup which is
responsible for the decay of the  first neutron.
\begin{figure}[ht]
\centering
\includegraphics[scale=0.5,angle=90]{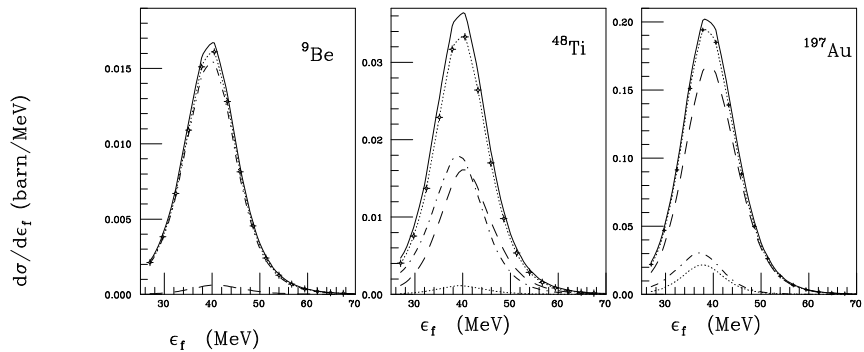}
\caption{Neutron energy distributions following breakup of $^{11}$Be at 41A.MeV for
several targets ($^{9}$Be, $^{48}$Ti, $^{197}$Au). Nuclear breakup distributions are
represented by the dot-dashed line, Coulomb by the long dashed line, the Coulomb-nuclear
coupling term by the dotted line, their  coherent sum by the solid line. The dotted
line with crosses is the nuclear plus the perturbative Coulomb incoherent sum. }
\end{figure}

We show in Figs.(3)
the neutron final energy spectra for  Coulomb and for  nuclear breakup  for the reactions
\noindent $^{9}Be(^{11}Be,^{10}Be+n)^{9}Be$, $^{48}Ti(^{11}Be,^{10}Be+n)^{48}Ti$,
$^{197}Au(^{11}Be,^{10}Be+n)^{197}Au$ at $E_{inc}$=41A.MeV \cite{anne} in the target reference
frame.
 We have used the same low energy
cutoff (27MeV) as in the experiment.  We see that the nuclear breakup
(dot-dashed line) is dominant for the $^{9}$Be target, for the $^{48}$Ti target Coulomb 
(long dashed line) and nuclear  give similar contributions, and
for $^{197}$Au, Coulomb breakup becomes dominant. The Coulomb-nuclear coupling term (dotted line) increases with the
target mass and for $^{197}$Au is of the same order as the pure nuclear term. 
From these results, we would expect big
interferences between Coulomb and nuclear processes for
$^{48}$Ti and $^{197}$Au. In fact 
the solid line which represents the coherent sum of the three processes is rather different from the dotted
line with crosses, which is the incoherent sum of nuclear and perturbative Coulomb
breakup.

  Finally in  Fig.(4a) we show the neutron energy spectrum in the
projectile reference frame together with the data from
\cite{nak} for Coulomb breakup on
$^{208}Pb$ at
$72$A.MeV. The same spectrum is also given in the target reference frame in Fig.(4b). Notation
is the same as in Fig.(1). In this case as in the case of the Au target discussed before the
contribution of the Coulomb-nuclear coupling term is very close to that of the pure nuclear
term. Our results agree well with the experimental spectrum. We used the known spectroscopic
factor
$C^2S=0.77$ \cite{as} for the s-initial state in $^{11}$Be. 
The integrated Coulomb cross section is 1.70b to be
compared with the experimental value of 1.8$\pm$ 0.4b. This is to show that our treatment of the Coulomb breakup
by perturbation theory is accurate within the experimental error bars for the description of the shape and
absolute magnitude of Coulomb breakup. It has been suggested \cite{kys,mb,jas,trj,nt}
that a treatment beyond perturbation theory
could be more accurate in same cases. Proton breakup, in particular at lower energies than those discussed here, is
one of such cases. On the other hand methods that have studied consistently
 neutron breakup \cite{tb,tf} in first order
or higher order, within the same model, have found negligible differences. Non perturbative
treatments are also expected to be more accurate in the case of smaller neutron
separation energies and lower beam energies \cite{trj} than those discussed here. The spectrum of Fig.(4) has been
calculated and discussed in a number of theoretical works \cite{tf,ban,cha} which all use
different non-perturbative methods but whose results are all consistent with each
other and with our results, within the experimental error bars. 
\begin{figure}[ht]
\centering
\includegraphics[scale=0.5,angle=90]{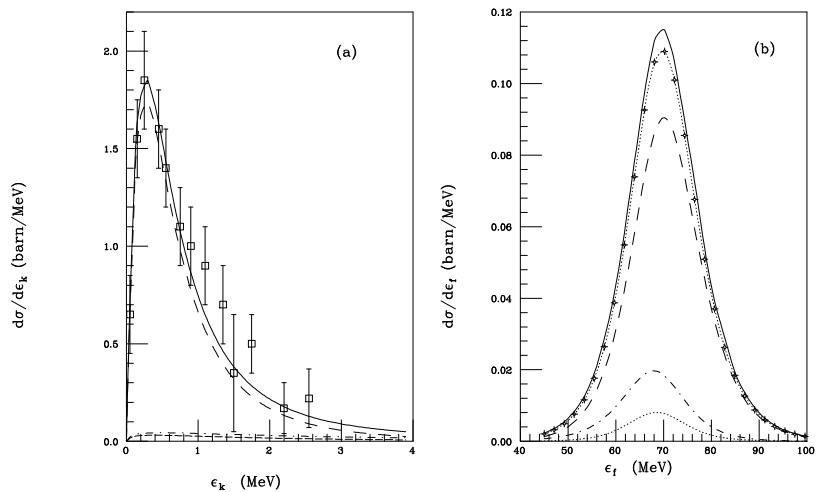}
\caption{ Neutron energy distributions following breakup of $^{11}$Be on $^{208}$Pb at
72A.MeV  in the projectile frame (a), and in the target frame (b). The notation is
the same as in Fig.(1). Experimental points are from Nakamura [4].}
\end{figure}

\subsection{Interference effects in the neutron angular distributions }
\begin{figure}[ht]
\centering
\includegraphics[scale=0.5,angle=90]{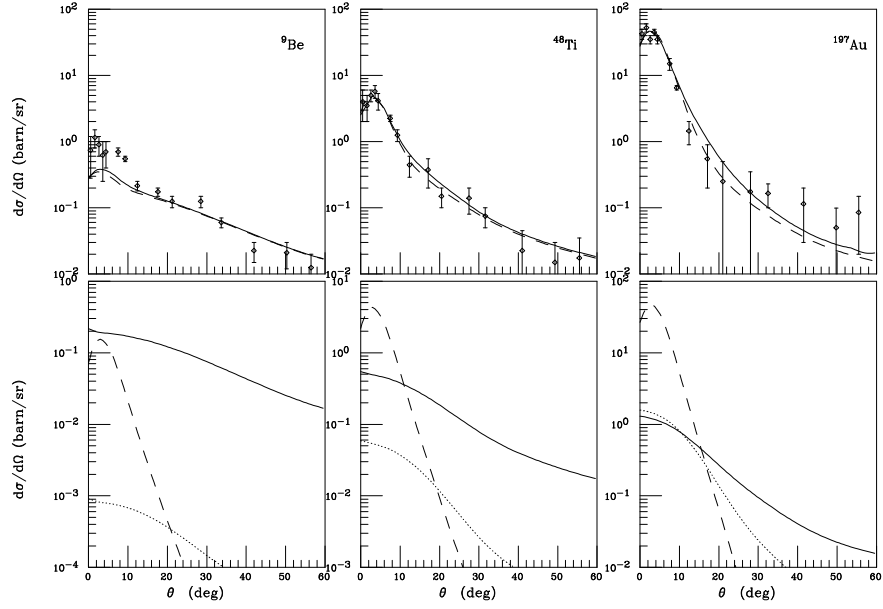}
\caption{Neutron angular distributions following breakup of $^{11}$Be at 41A.MeV  for
several targets ($^{9}$Be, $^{48}$Ti, $^{197}$Au). In the bottom figures the nuclear distribution is
represented by the solid line, Coulomb by the long dashed line, and the Coulomb-nuclear term by the dotted line.
In the top figures the
coherent sum of the nuclear, Coulomb and Coulomb-nuclear terms by the solid line, the nuclear plus the
perturbative Coulomb incoherent sum by the 
dashed line. Experimental points are from Anne \cite{anne}.}
\end{figure}

Fig.(5) contains the data from the exclusive experiment of Ref.
\cite{anne} in which the neutron angular distribution following the breakup from $^{11}$Be
was measured in coincidence with the $^{10}$Be ejectile. Three different targets were used
to check on the relative importance of the nuclear and Coulomb breakup mechanisms.
 In the bottom part of the figure
 the solid line is the nuclear breakup cross section, the long dashed line is the
Coulomb breakup while the dotted line is the Coulomb-nuclear coupling term.  In the top part of the figure we show
the experimental data together with the coherent sum of the three contributions (solid line). The dashed line is
the nuclear plus the perturbative Coulomb incoherent sum. It is important to notice that the large angle
scattering is due mainly to the nuclear breakup and it is sensitive to the neutron-target optical
potential used. About two thirds of the  total amount of nuclear breakup is due to large angle
scattering. Therefore new measurements of neutron angular distributions, including large angle scattering
data would greatly help in settling the question of the relative importance of nuclear and Coulomb
breakup for heavy targets. On the other hand spectra of the type shown in Fig.(4a) correspond to small
angle scattering only (typically $\theta<15$deg) where the nuclear is very small and therefore fitting
them is not too significative for the understanding of the relative importance of the two reaction
mechanisms.

\begin{figure}[ht]
\centering
\includegraphics[scale=0.5,angle=90]{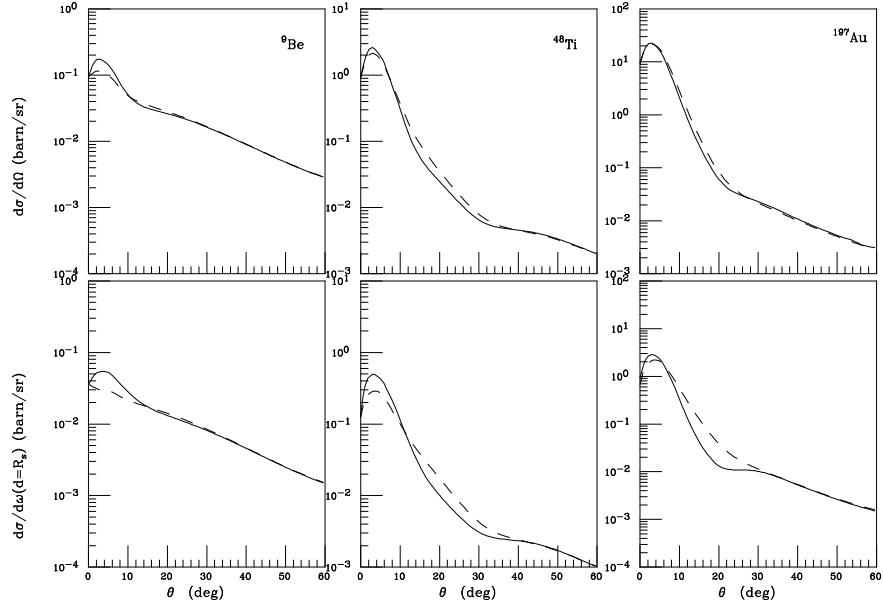}
\caption{Neutron angular distribution as in Fig.5, but 0$<\phi<$30 degrees for
several targets ($^{9}$Be, $^{48}$Ti, $^{197}$Au). The results in the top figures are integrated over the core
target impact parameter d. The bottom figures are results at a fixed impact parameter $d=R_s$. The  solid
line represents the coherent sum of nuclear, Coulomb and Coulomb-nuclear terms. Their incoherent sum by the dashed
line.}
\end{figure}

 The theoretical results for the neutron angular distributions were summed over the neutron
energy spectrum and averaged over its out-of-plane angle
$\phi$ because this is what is contained in the experimental data. The peak values of the cross section show
clearly an increase with the target due to the Coulomb breakup. From the comparison of  the coherent and incoherent
sums it is clear that coupling and interference effects are more important at the angles where nuclear and Coulomb
terms are closer in magnitude, which happens at small $\theta$ for the $^{9}$Be target  and around 20-30deg for the
other two targets.

We have seen that the effects due to the interference
are partially washed out by the average over
$\phi$. On the other hand we have identified the reason of the lost of coherence at large $\theta$ in the integration
of the probabilities over the core-target impact parameter d as contained in Eq.(\ref{csec}).
To clarify these points we show in Fig. (6)  similar calculations for the same
targets as before but this time we have chosen to restrict the $\phi$ integration to the range $0\div 30$deg. In
the top part the integral over the core-target impact parameter has been performed. In the bottom part of the figure
the calculations have been performed integrating one fermi out of the  strong absorption radius $R_s$.
The results of these figures are however still summed over the neutron energy spectrum. 
\begin{figure}[ht]
\centering
\includegraphics[scale=0.5,angle=90]{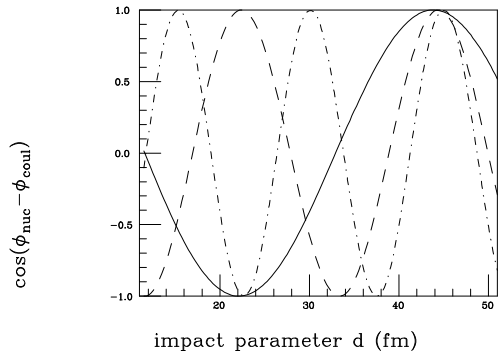}
\caption{Cosinus of the Coulomb-nuclear phase as a function of the impact parameter
for several $\theta$ angle and $\varepsilon_f=41MeV$ (fixed $k_z$, $k_y=0$, $k_{\perp}=k_x$).
The solid line stands for $\theta=6$ degrees, the dashed line for
$\theta=12$ degrees and the dot-dashed line for
$\theta=18$ degrees. The target is $^{197}$Au.}
\end{figure}

The solid lines are the results for the cross section obtained by summing coherently the
nuclear and Coulomb and Coulomb-nuclear amplitudes while the dashed line is again the
incoherent sum of the nuclear plus the perturbative Coulomb. From the bottom figure it is clear
that the interferences change sign as a function of $\theta$.

Eqs.(\ref{veff2}) and (\ref{amp5}) show that the important contributions to the integral for
the perturbative Coulomb amplitude come from the projectile region while
integral (\ref{amp3}) for the nuclear amplitude is weighted towards the target.  As a
consequence the Coulomb-nuclear phase difference has a dependence on the
projectile-target impact parameter. The cosine of the relative
Coulomb-nuclear phase as a function of $d$ is shown in Fig.(7) for  a set of
neutron emission angles. The neutron is emitted in the plane defined by the
impact parameter and the incident beam direction.  The cosine of the relative
phase has a regular oscillatory behavior and the period of the oscillations
decreases with  increasing $\theta$. The oscillations reduce the magnitude of
the Coulomb-nuclear interference. They also influence its sign as indicated
by the integrated cross sections in Table 2.

 \begin{table}
\caption {Integrated ($\theta=0\div 60$deg) breakup cross sections in barn. Incident
energies in A.MeV. }
\begin{center}
\begin{tabular}{|cc|cccccc|}\hline
{\footnotesize Target}& E$_{inc}$  & 
Coul-Nucl&Nuclear& Coulomb& Incoh. sum&Cohe. sum&\%\\
\hline \hline
$^{9}$Be &30 & 0.6 10$^{-3}$ & 0.200 & 0.011 &0.212 & 0.222&5 \\
        &41 & 0.5 10$^{-3}$ & 0.191& 0.009& 0.200 & 0.209& 5 \\
       & 120 & 0.4 10$^{-4}$ & 0.044 & 0.004 & 0.048 & 0.050 &5  \\ \hline
$^{48}$Ti&30 &0.2 10$^{-1}$ & 0.260 & 0.292&0.573& 0.621&9\\
         & 41 & 0.2 10$^{-1}$& 0.252 & 0.229 & 0.496&0.531&8\\
         &120 & 0.6 10$^{-2}$& 0.234 & 0.098& 0.338&0.374&10\\ \hline
$^{197}$Au&30 & 0.327 & 0.372 & 2.98& 3.68& 3.50&-5\\ 
          &41 & 0.245 & 0.354 & 2.36& 2.96 &2.85&-4 \\
          &120& 0.151 & 0.321 & 1.05& 1.52 &1.53&1\\ \hline   
$^{208}$Pb&72 & 0.148 & 0.346 & 1.70& 2.19& 2.17&-1\\\hline
\end{tabular}
\end{center}
\end {table}

  Finally
we give in Table 2 the absolute cross sections for each individual term: the
Coulomb-nuclear amplitude,  the eikonal nuclear and Coulomb perturbative. Also their
coherent and incoherent sums are shown, for the targets studied in this work.
We have integrated in the experimental range $\theta=0\div 60$deg. In the
calculations of this work we have used the strong absorption hypothesis
Eq.(\ref{prs1}) with the prescription
$R_s=1.4 (A_P^{1/3}+A_T^{1/3})fm$. Using the smooth absorption hypothesis \cite{27} would reduce the absolute
nuclear cross section
 but leave the energy and momentum distributions unchanged \cite{me}. For example the
total coherent cross sections on $^{9}$Be at the three given energies, with a smooth cut
off are 0.221b, 0.200b and 0.042 b respectively. The same would happen using the new
formalism proposed in
\cite{BB1}. The values without interference are very similar to those already obtained by
some of us in
\cite{bbn}. We remind the reader that when the neutron is detected in coincidence with
the core only the diffraction-type of nuclear breakup contributes. Cross section values
are given at three incident energies
$E_{inc}=30, 41 $ and $120$A.MeV and at $72$A.MeV for the $^{208}$Pb target. In the last column we give the
percentage of interference effect. The effect of the interference on the angle integrated energy distribution is of
the order of $4\div 10\%$ therefore small as already seen by other authors \cite{ts,jas,BB}. The interference is
constructive for the $^{9}$Be and $^{48}$Ti targets but destructive  for the $^{197}$Au  and
$^{208}$Pb targets. For the heavy targets it appears also that the interferences tend to vanish at high
energy. Then from the analysis of the energy and angular
distributions we conclude that the effect of the new Coulomb-nuclear coupling  term and of the interferences is to
give a slight increase with respect to the simple sum of the nuclear plus the Coulomb perturbative breakup. However
from the analysis of the results for the total cross sections it appears that the total interference of the three
terms can be destructive with respect to their sum for the heavy target. This is manly due to the imaginary part of
the neutron target optical potential.  The situation might be different in the case of the proton breakup studied in
\cite{dlv,st,nt,eh} because  the  effective charge in the Coulomb potential Eq.(\ref{a1}) will be different from
ours obtained   for a neutron.

At this stage it is difficult to draw a general conclusion on the effects of the
interference and of the Coulomb-nuclear coupling as they clearly depend on the target
and on the incident energy and on the angular range covered by the experiment. 

The results discussed in this paper are restricted to the 2s initial state in
$^{11}$Be. The Coulomb breakup from the d-component of the ground state gives a
contribution two orders of magnitude smaller than the s-component and thus it has
been neglected here. However the formulae given in the appendix hold for any initial
angular momentum. Calculations for cases like the weakly bound carbon isotopes, where
$l_i=2$ orbitals are important, will be presented elsewhere.

\section{Conclusions}

In this work  we have derived an eikonal formalism leading to  Eq.(\ref{amp1}) to treat consistently the nuclear
and Coulomb breakup to all orders in the interactions. This formula could be studied numerically. Here we have
studied only the effect of the first order coupling and interference between the two processes.  By taking first
order terms in the Coulomb potential and the nuclear potential to all orders we have  obtained a scheme in which the
breakup amplitude is a sum of three terms.
 Two of
them reduce to the well known forms of the nuclear breakup in the eikonal model and of the
Coulomb breakup in  first order perturbation theory. The third term, Eq.(\ref{amp6}) which
is derived here for the first time, depends on both the Coulomb and nuclear potentials and
can be viewed as one of  the lowest "higher order coupling terms". Within such a model
numerical results  show that interferences between nuclear and Coulomb breakup are present
in halo nucleus reactions and in the data already existing. They are responsible for the
peak at small angles in the angular distribution on the $^{9}$Be target and for the
behavior at large $\theta$ in the case of heavier targets. They have not been seen so far
more clearly in the data because of the averaging over the neutron out-of-the-reaction
plane angle
$\phi$ and the integration over the core-target impact parameter. However they would show up  in
  neutron double differential angular distribution from halo breakup by making a
selection  on a small range of
$\phi$ angles and detecting the core at fixed angles, provided it follows a Coulomb trajectory. In this case the
reaction plane could be identified and a one by one correspondence between the core detection angle and the
trajectory impact parameter could be applied.

The Coulomb-nuclear interference effects are small for the cases studied in
the this paper. 
Interference terms do not have a definite sign in the angular distribution but they are
overall constructive  in the energy spectra
and absolute cross sections at medium energies and for light to medium mass targets. For
heavy targets the results presented here show an overall destructive interference which depends on the imaginary
part of the neutron target optical potential and which tends to vanish increasing the beam energy. The effect of the
new Coulomb-nuclear coupling term is rather small and negligible for light targets, while for heavy targets is of the
same order as the pure nuclear term. This is evident in the energy distributions and in the small angle angular
distributions.  The comparison with the data and with other works is reassuring in so far as
the neglect of higher order terms in the Coulomb amplitude is  concerned.
They scale as powers of the effective charge, that is as powers of
$C=\beta_{1}Z_{P}Z_{T}e^{2}/\hbar v$ and $\beta_{1}$ is small ($\beta
_{1}=0.1)$  for the $^{11}$Be projectile.  Higher order terms will be very
small for light targets but could become more important for heavy targets.
They have not been studied here but they can  be  evaluated   within the adiabatic approximation
Eq.(\ref{sudd}). 

  Thus it appears that the so-called Coulomb experiments  would need a more
accurate theoretical analysis than performed so far. Also we think that it would be very important to measure more
neutron angular distributions, like the ones
\cite{anne} discussed in this paper. They are a beautiful tool to identify the reaction
mechanism and to test the theoretical methods.
Therefore we suggest that
in more exclusive kind of experiments it should be possible to see better the effects discussed in
this work. It will be necessary to have more intense radioactive beams giving 
 also a better statistics and to 
use neutron position sensitive detectors over a wide angular range. New facilities of the
type being planned in Europe (EURISOL) and USA (RIA) are therefore most welcome.

{\bf Acknowledgments}

We wish to thank E. Piasecki for drawing Ref.\cite{cha1} to our attention and for suggesting Fig.(2).

\appendix\section{Details of the Coulomb and Coulomb-nuclear amplitudes.}

The Coulomb contribution to the effective interaction is%
\begin{equation}
V_{eff}(\mathbf{r},t)=+{\beta}_{1}Z_{P}Z_{T}e^{2}\frac{xd+zvt}{\left(
d^{2}+\left(  vt\right)  ^{2}\right)  ^{3/2}} \label{a1}%
\end{equation}
We need to evaluate several integrals. The first is the Fourier transform%

\begin{equation}
\chi_{eff}\left(  \mathbf{r},\omega\right)  =\frac{1}{\hbar}\int_{-\infty
}^{\infty}dte^{i\omega t}V_{eff}\left(  \mathbf{r},t\right)  \label{a2}%
\end{equation}
By making a change of variables $s=vt/d$ the first integral becomes

\begin{eqnarray}
\chi_{eff}\left(  \mathbf{r},\omega\right)   &  =&\frac{C}{d}\int_{-\infty
}^{\infty}dse^{i\varpi s}\frac{x+zs}{\left(  1+s^{2}\right)  ^{3/2}
}\nonumber \\
&  =&2\frac{C}{d}\left(  x\varpi K_{1}\left(  \varpi\right)  +iz\varpi
K_{0}\left(  \varpi\right)  \right)  \label{a3}
\end{eqnarray}
where
\begin{equation}
\varpi=\frac{\omega d}{v}=\frac{(\varepsilon_{\mathbf{k}}-\varepsilon_{0}%
)d}{\hbar v}.\label{ad1}%
\end{equation}
and the constant $C={\beta}_{1}Z_{P}Z_{T}e^{2}/\hbar v$ is a dimensionless
interaction strength. The functions$K_{0}$ and $K_{1}$ are modified Bessel
functions and can be defined by the integrals%
\begin{eqnarray}
\varpi K_{1}\left(  \varpi\right)   &  =&{\frac{1}{2}}\int_{-\infty}^{\infty
}\frac{\cos\left(  \varpi t\right)  dt}{\left(  1+t^{2}\right)  ^{3/2}
}\label{K1}\\
\varpi K_{0}\left(  \varpi\right)   &  =&{\frac{1}{2}}\int_{-\infty}^{\infty
}\frac{t\sin\left(  \varpi t\right)  dt}{\left(  1+t^{2}\right)  ^{3/2}
}\label{K0}
\end{eqnarray}
Substituting eq.(\ref{a3}) into eq.(\ref{amp5}) yields the perturbation
amplitude
\begin{eqnarray}
g_{r}^{pert}\left(  \mathbf{k,}\mathbf{d}\right)   &  =&-i\frac{2C}{d}\int
d^{3}\mathbf{r}e^{-i\mathbf{k.r}}\left(  x\varpi K_{1}\left(  \varpi\right)
+iz\varpi K_{0}\left(  \varpi\right)  \right)  \phi_{lm}\left(  \mathbf{r}
\right) \nonumber \\
&  =&2\frac{{C}}{d}\left(  \varpi K_{1}\left(  \varpi\right)  \frac{d}{dk_{x}
}+i\varpi K_{0}\left(  \varpi\right)  \frac{d}{dk_{z}}\right)  \tilde{\phi
}_{lm}\left(  \mathbf{k}\right) \label{amp8} .
\end{eqnarray}

The constant $C={\beta}_{1}Z_{P}Z_{T}e^{2}/\hbar v$ is a dimensionless
interaction strength and $\varpi$\ is the adiabaticity parameter. The
functions $K_{0}$ and $K_{1}$ are modified Bessel functions and $\tilde{\phi
}_{lm}\left(  \mathbf{k}\right)  $ is the Fourier transform of $\phi
_{lm}\left(  \mathbf{r}\right)  $. First order perturbation theory in
$\ V_{eff}$ should be a good approximation if $CR_{p}/d < 1/2$ where
$R_{p}$ is the radius of the halo nucleus. Numerical values for the modified
Bessel functions show that the adiabatic or 'frozen halo' approximation should
be satisfactory if $\varpi < 0.2$. \ Eq.(\ref{amp8}) is the standard
dipole expression for Coulomb breakup.

The function

\begin{equation}
B\left(  x,z,\varpi\right)  =\frac{i}{\hbar}\left[  \int_{-\infty}%
^{0}dte^{i\omega t}V_{eff}\left(  \mathbf{r},t\right)  +\int_{0}^{\infty
}dtV_{eff}\left(  \mathbf{r},t\right)  \right]
\end{equation}
involves two integrals. The first is%
\begin{equation}
\frac{1}{\hbar}\int_{0}^{\infty}dtV_{eff}\left(  \mathbf{r},t\right)
=\frac{C}{d}\left(  x+z\right)
\end{equation}
The other is a one-sided Fourier transform%
\begin{equation}
\frac{1}{\hbar}\int_{-\infty}^{0}dte^{i\omega t}V_{eff}\left(  \mathbf{r}%
,t\right)  =\frac{1}{\hbar}\int_{0}^{\infty}dte^{-i\omega t}V_{eff}\left(
\mathbf{r},-t\right)
\end{equation}
It can be expressed in terms of $\varpi K_{0}\left(  \varpi\right)  ,\varpi
K_{1}\left(  \varpi\right)  $ and two new integrals%

\begin{eqnarray}
\varpi\bar{K}_{1}\left(  \varpi\right)   &  =&\int_{0}^{\infty}\frac
{\sin\left(  \varpi t\right)  dt}{\left(  1+t^{2}\right)  ^{3/2}}
\label{Kbar1}\\
\varpi\bar{K}_{0}\left(  \varpi\right)   &  =&\int_{0}^{\infty}\frac
{t\cos\left(  \varpi t\right)  dt}{\left(  1+t^{2}\right)  ^{3/2}
}\label{Kbar0}
\end{eqnarray}
Putting everything together%
\begin{equation}
B\left(  x,z,\varpi\right)  =i\frac{C}{d}\left(  x(\varpi K_{1}\left(
\varpi\right)  +1-i\varpi\bar{K}_{1}\left(  \varpi\right)  )-z\left(
\varpi\bar{K}_{0}\left(  \varpi\right)  -1-i\varpi K_{0}\left(  \varpi\right)
\right)  \right)
\end{equation}

\section {\bf Initial states with general angular momenta.}
\subsection {The Coulomb breakup amplitude.}

If the initial state wave function is approximated by its asymptotic form which is an
Hankel function \cite{sh}

   \begin{equation}\phi_{lm}(r)=-i^lC_i\gamma_0h_{l}^{(1)}(i\gamma_0r)
   Y_{lm}(\theta,\phi), \,\,\,\gamma_0 r >>1,
   \label{in}
   \end{equation} 
where $C_i$ is the asymptotic normalization constant
\cite{bbp} and $\gamma_0=\sqrt{-2m\varepsilon_0}/\hbar$, 
then the general
form of the initial state momentum distribution is given by the Fourier transform of
Eq.(\ref{in})
 \begin{equation} \widetilde {\phi}_{lm}({\bf k_{\perp}},k_z) = 4\pi  C_1 {{k}^l\over
\gamma_0^l(k_{\perp}^2+k_z^2+\gamma_0^2)}Y_{l,m} (\hat{k}) 
 \label{2}  \end{equation} where ${\bf {k}}\equiv (k_x,k_y,k_z)$ is a real vector.

In nuclear breakup reactions the use of the asymptotic part of the
initial bound state wave function is justified because of the short
range nature of the nuclear interaction which gives breakup form factors
in Eq.(\ref{amp3}) localized in the overlap region between the two
interacting nuclei.  The
amplitude for Coulomb breakup is expressed in terms of the Fourier
transform of the bound state wave function in Eq.(\ref{amp8}). This Fourier
transform is well approximated by the Fourier transform Eq.(\ref{2}) of
the corresponding Hankel function provided that $\gamma_0 R << 1$ and
$kR << 1$ where $R$ is the nuclear radius.  These conditions are
often satisfied in the Coulomb breakup of a halo nucleus.  We have
checked this for $^{11}$Be using wave functions calculated in a
square-well potential and have shown that for
$|\varepsilon_0|\leq 1$MeV and up to $l=2$ the contribution of the
internal part of the wave function to the full Fourier transform can be
neglected.

We want to evaluate $2 \pi <\sum_m |A_{lm}^{Coul}|^2>_\varphi$ where $<>_\varphi$
stands for the average over the $\varphi$ angle of ${\bf k}_f$.
Some definitions for the neutron momenta are: 
${\bf k}_f=(k_x,k_y,k_{f_z})$ is the neutron final momentum in the target reference frame,
while 
${\bf k}=(k_x,k_y,k_z)$ is in the projectile reference frame and $k_z=k_{f_z}-m_nv/\hbar$,
$k_\perp^2=k_x^2+k_y^2$. 

We define
\begin{equation}
{\cal A}_{lm}^{Coul}=\left( K_1(\bar \omega)\frac{\partial }{\partial k_{x}}+
iK_0(\bar \omega)\frac{\partial }{\partial k_z}\right)
F_{lm}({\bf k})\label{aa1}
\end{equation}
where $K_1$ and $K_2$ are  modified Bessel functions and
\begin{eqnarray}
F_{lm}(\hat{k})&=& f_l(k) Y_{lm}(\hat{\bf k}) \\
f_l(k)&=&\frac{k^l}{\gamma_0^l(\gamma_0^2+k_\perp^2+k_z^2)}
\end{eqnarray}
with $\gamma_0=\sqrt{-2m\epsilon_0}/h$. Then
\begin{equation}
{\cal B}_l=\sum_m F_{lm}^*({k}')F_{lm}({k})=\frac{2l+1}{4\pi}f(k')f(k)P_
l(\cos\Theta)
\end{equation}
with $\cos\Theta={\bf k}.{\bf k}'/k'k$.

We calculate the derivatives like $\partial^2{\cal B}_l/\partial k'_x
\partial k_x $ and finally put ${\bf k}'={\bf k}$.
Note that $\partial/\partial k'_x P_l(\cos\theta)
|_{{\bf k}={\bf k}'}=0$.

Using the above results in Eq.(\ref{aa1})  and averaging over $\varphi$
\begin{eqnarray}
2 \pi <\sum_m |{\cal A}_{lm}^{Coul}|^2>_\varphi&=&\frac{2l+1}
{2 (k_\perp^2+k_z^2)^2}
\left(K_1^2(\bar \omega)\left[f'^2 k^2 k_\perp^2/2 + P'_l(1)f^2(k_z^2+k_\perp/2)
\right] \right. \nonumber \\
&+&\left. K_0^2(\bar \omega) \left[f'^2 k^2 k_z^2/2 + P'_l(1) f^2 k_\perp^2
\right] \right)
\end{eqnarray}

After simplifications, the final result is
\begin{eqnarray}
 2 \pi <\sum_m |{\cal A}_{lm}^{Coul}|^2>_\varphi&=&\frac{2l+1}{2}
\left( K_1^2\, [{\cal U}_l + \frac{1}{2}k_\perp^2 {\cal V}_l]
+ K_0^2\, [{\cal U}_l + k_z^2 {\cal V}_l]
\right)
\end{eqnarray}
with
\begin{eqnarray}
{\cal U}_l(k) &=& \frac{l(l+1)}{2} \frac{k^{2l-2}}
{\gamma^{2l}(\gamma_0^2+k^2)^2} \\
{\cal V}_l(k) &=& \frac{k^{2l-4}}{\gamma_0^{2l}(\gamma_0^2+k^2)^4}
\left[ \frac{l(l-1)}{2}(\gamma_0^2+k^2)^2-4lk^2(\gamma_0^2+k^2)+4k^4
\right]
\end{eqnarray}

\subsection {The Coulomb cross section.}

We give in the following  the analytical formulae for the cross section due to Coulomb
breakup only obtained using  Eqs.(\ref{csec},\ref{prs1},\ref{amp8}). In the integral
Eq.(\ref{csec}) over the impact parameter d we change variable of integration from d to
$\bar
\omega$ given in Eq.(\ref{ad1}). Furthermore to compare to experimental angular
distributions we average over the angle
$\phi$. The expressions for the double
differential cross section, in the case of an initial s and p
state are 
\begin{equation}{d^2\sigma_{Coul}\over d\varepsilon_fd\Omega}={mk_f\over
\hbar^2} \left ({C C_1\bar{\bar \omega}}\right
)^2{2\over \pi (k_{\perp}^2+k_z^2 +\gamma_0^2)^4}\left[
{\cal A }(K_1^2-K_0^2)+{\cal B}K_0K_1\right ]\label{dc}\end{equation}
where
\begin{equation}{\cal A}= 4\left(k_z^2-{k_{\perp}^2\over
2})\right)\end{equation} and
\begin{equation}{\cal B}=4{k_{\perp}^2\over\bar{\bar
\omega}}\end{equation} in the case of an s-initial state, while
for a p-initial state we get 

\begin{equation}{\cal A}={1\over
\gamma_0^2}\left (k_{\perp}^4+(k_z^2+\gamma_0^2)^2+{k_{\perp}^2\over
2(k_{\perp}^2+\gamma_0^2)}(k_{\perp}^2+\gamma_0^2-k_z^2)^2
+4k_z^2(2k_{\perp}^2+\gamma_0^2)\right)\end{equation}

\begin{equation}{\cal B}=-{1\over \gamma_0^2}\left ({2\over \bar{\bar \omega}}
\left(k_{\perp}^4+(k_z^2+\gamma_0^2)^2+{k_{\perp}^2\over
2(k_{\perp}^2+\gamma_0^2)}(k_{\perp}^2+\gamma_0^2-k_z^2)^2\right)\right)\end{equation}
$\bar{\bar \omega}={\bar \omega}(d=R_s)$ is the argument of the modified Bessel functions in
Eq.(\ref{dc}).

\end{document}